\documentclass[aps,prc,preprint,showpacs,superscriptaddress,amsmath,amssymb,floatfix]{revtex4}

\usepackage[dvips]{graphicx,color}
\usepackage{dcolumn}
\usepackage{bm}

\def\pmb#1{\setbox0=\hbox{#1}%
  \kern-.025em\copy0\kern-\wd0 
  \kern.05em\copy0\kern-\wd0
  \kern-.025em\raise.0433em\box0 }

\newcommand{\Figuretable}[1]{%
  \begin{center} --------- {\bf #1} --------- \\ \end{center}} 
\def\lambdabar{\protect\@lambdabar}
\def\@lambdabar{%
\relax
\bgroup
\def\@tempa{\hbox{\raise.73\ht0
\hbox to0pt{\kern.25\wd0\vrule width.5\wd0
height.1pt depth.1pt\hss}\box0}}%
\mathchoice{\setbox0\hbox{$\displaystyle\lambda$}\@tempa}%
{\setbox0\hbox{$\textstyle\lambda$}\@tempa}%
{\setbox0\hbox{$\scriptstyle\lambda$}\@tempa}%
{\setbox0\hbox{$\scriptscriptstyle\lambda$}\@tempa}%
\egroup
}

\begin{document}

\preprint{J-PARC-TH-0006}

\title{\boldmath
Angular distributions of elastic scattering of $\Sigma^-$ hyperons from nuclei
and the $\Sigma$-nucleus potentials 
}

\author{Toru~Harada}%
\email{harada@isc.osakac.ac.jp}
\affiliation{%
Research Center for Physics and Mathematics,
Osaka Electro-Communication University, Neyagawa, Osaka, 572-8530, Japan
}
\affiliation{%
J-PARC Branch, KEK Theory Center, Institute of Particle and Nuclear Studies,
High Energy Accelerator Research Organization (KEK),
203-1, Shirakata, Tokai, Ibaraki, 319-1106, Japan
}

\author{Yoshiharu~Hirabayashi}%
\affiliation{%
Information Initiative Center, 
Hokkaido University, Sapporo, 060-0811, Japan
}

\date{\today}

\begin{abstract}
We theoretically investigate the elastic scattering of 50-MeV $\Sigma^-$ hyperons
from $^{28}$Si and $^{208}$Pb in order to clarify the radial distribution 
of $\Sigma$-nucleus (optical) potentials.
The angular distributions of differential cross sections 
are calculated using several potentials that can explain experimental data 
of the $\Sigma^-$ atomic X-ray and ($\pi^-$, $K^+$) reaction spectra 
simultaneously. 
The magnitude and oscillation pattern of the angular distributions 
are understood by the use of  nearside/farside decompositions of 
their scattering amplitudes. 
It is shown that the resultant angular distributions 
provide a clue to discriminating among the radial distributions of 
the potentials that have a repulsion inside the nuclear surface 
and an attraction outside the nucleus 
with a sizable absorption. 

\end{abstract}
\maketitle


\section{Introduction}

One of the fundamental subjects in hypernuclear study is 
to understand the properties of hyperon-nucleus interactions.
Thus it has been discussed that  a study of a negatively charged $\Sigma^-$ 
hyperon in nuclear medium would provide valuable information concerning 
the maximal mass of neutron stars,
in which a baryon fraction is found to depend on properties of 
hypernuclear potentials for neutron stars in astrophysics
 \cite{Balberg97,Baldo00,Takatsuka01,Schaffner10}.

The systematic study of the $\Sigma$-nucleus (optical) potential 
based on the $\Sigma^-$ atomic X-ray data was performed 
by Batty and his collaborators 
\cite{Batty78,Batty94,Batty97}.  
The latest analyses of the $\Sigma^-$ atomic X-ray have suggested that 
the $\Sigma$-nucleus potential has a repulsion 
inside the nuclear surface and a shallow attraction outside 
the nucleus with a sizable absorption \cite{
Batty94,Yamada94,Mares95,Harada95,Dabrowski02}. 
Noumi and his collaborators \cite{Noumi02,Saha04} have performed 
measurements of $\Sigma$-hypernuclei by inclusive ($\pi^-$, $K^+$) reactions 
on C, Si, Ni, In and Bi targets at $p_{\pi}$=1.20 GeV/c in KEK-E438 experiments. 
Their analyses of the $\Sigma^-$ quasi-free (QF) spectra 
with a distorted-wave impulse approximation found that the 
$\Sigma$-nucleus potentials have a strong repulsion in the real part 
and a sizable absorption in the imaginary part within a Woods-Saxon (WS) form:
\begin{equation}
U_\Sigma(r)= (V_0^\Sigma + i W_0^\Sigma)/[1 + \exp{((r-R)/a)}],
\label{eqn:e1}
\end{equation}
where ($V^\Sigma_0$, $W^\Sigma_0$)=($+90$ MeV, $-40$ MeV) 
with $R=$ $1.1A_{\rm core}^{1/3}$ and $a=$ 0.67 fm \cite{Saha04}.

In previous papers \cite{Harada05,Harada06}, we have succeeded 
to explain simultaneously the data of the $\Sigma^-$ atoms and 
the ($\pi^-$, $K^+$) reactions on $^{28}$Si and $^{208}$Bi, using 
the $\Sigma$-nucleus potentials that have a repulsion inside the nuclear surface 
and an attraction outside the nucleus with a sizable absorption.
This repulsion originates from the $\Sigma N$ $T=$ 3/2, $^{3}S_1$ 
channel \cite{Dover89,Harada90,Harada91},
whose state corresponds to a quark Pauli-forbidden state 
in the baryon-baryon system \cite{Oka87,Fujiwara07,Rijken08},
and it is a candidate for the appearance of quark degrees of freedom in nuclear physics.
Theoretical analyses of the ($\pi^-$, $K^+$) reaction \cite{Harada05,Harada06} 
distinguish partially among properties of the $\Sigma$-nucleus potential 
that can reproduce the $\Sigma^-$ atomic X-ray data sufficiently, 
whereas the radial distribution of the $\Sigma$-nucleus potential inside the nucleus and 
its strength at the center are hardly determined by fits to the $\Sigma^-$ QF spectrum. 
Moreover, we have recognized that an energy dependence of $({d \sigma/d \Omega})^{\rm opt}$ 
in elementary $\pi^- + p \to K^+ + \Sigma^-$ processes in nuclei is needed 
to explain the behavior of the ($\pi^-$, $K^+$) spectrum \cite{Harada05}.
Even if using near-recoilless ($K^-$, $\pi^+$) reactions, 
the radial distribution of the $\Sigma$-nucleus potential cannot clearly be 
determined for a suitable nuclear target such as $^{58}$Ni \cite{Harada09}.

Recently, Miwa and his collaborators~\cite{MiwaP40} have proposed an experiment
to measure scattering cross sections with high statistics
in $\Sigma^\mp p$ elastic and $\Sigma^-p \to \Lambda n$ inelastic scatterings 
by 500-MeV/c $\Sigma^\mp$ beam at J-PARC.
The purpose of this experiment is to test baryon-baryon interactions
based on the flavor SU(3) symmetry and to directly confirm the existence of 
the quark Pauli-forbidden state in baryon-baryon systems \cite{Oka87,Fujiwara07}. 

In this paper, we theoretically investigate the elastic scattering 
of $\Sigma^-$ hyperons from
nuclei in order to clarify the radial distribution of the $\Sigma$-nucleus 
(optical) potential.
We calculate the angular distributions of the differential cross sections   
in the elastic scattering of $\Sigma^-$ hyperons from $^{28}$Si 
and $^{208}$Pb at $E_{\rm lab}=$ 50 MeV,
and demonstrate the sensitivity of the angular distribution
to the radial distribution of several potentials 
that have explained the $\Sigma^-$ atomic X-ray data 
and the ($\pi^-$, $K^+$) spectra. 
It is well-known that optical potential models can describe the elastic 
scattering of protons or light ions from nuclei, 
employing appropriate potential parameters phenomenologically. 
The angular distribution provides specific tests of the validity 
of the optical potential,  
in comparison with the experimental data \cite{Watson69}.
This is a standard and promising approach for examining the radial 
distribution of the optical potential in nuclear physics, 
whereas some ambiguities may still be remained with their strong absorption.
Therefore, we believe to examine the radial distribution of the 
$\Sigma$-nucleus potentials with 
the elastic scattering of $\Sigma^-$ hyperons from nuclei.

\section{$\Sigma$-nucleus potentials}
\label{sect:pot}

Several theoretical attempts have been performed to construct 
a $\Sigma$-nucleus potential, 
fitting systematically to strong-interaction shifts and widths 
of $\Sigma^-$ atomic X-ray data
\cite{Batty94,Batty97}
and manifesting inclusive $K^+$ spectra in the ($\pi^-$,~$K^+$) reaction 
on nuclear targets \cite{Noumi02,Saha04,Kohno04,Harada05,Harada06}. 
The recent status of  understanding of the $\Sigma$-nucleus potential has been 
reviewed in Ref.~\cite{Friedman07}.
Here we briefly mention the $\Sigma$-nucleus potentials 
for $^{28}$Si and $^{208}$Pb that we used in this article. 
The detailed discussion on properties of the potentials are shown in 
Refs.\cite{Harada05,Harada06,Harada09}.

In previous papers \cite{Harada05,Harada06}, we have presented 
several types of the $\Sigma$-nucleus potential
obtained by fitting to strong-interaction shifts and 
widths of $\Sigma^-$ atomic X-ray for various nuclei. 
The $\Sigma$-nucleus potentials that we used are 
(a) the density-dependent (DD) potential \cite{Batty94},
(b) the relativistic mean-field (RMF) potential \cite{Mares95},
(c) the local-density approximation potential (LDA-NF) based on 
YNG-NF interaction \cite{Yamamoto85,Dabrowski02},
(d) the LDA potential (LDA-S3) based on phenomenological 
two-body $\Sigma N$ SAP-3 interaction \cite{Harada95},
(e) the shallow potential in the WS form (WS-sh) \cite{Hayano88}, 
and (f) the $t_{\rm eff}\rho$-type potential ($t_{\rm eff}\rho$)
\cite{Batty94}.
In Fig.~\ref{fig:1}, we display the real and imaginary parts of 
several $\Sigma$-nucleus potentials for $^{28}$Si, of which all
reproduce the experimental shifts and widths of the $\Sigma^-$ 
atomic $4f$ and $5g$ states sufficiently \cite{Harada05}.
The potentials for DD, RMF and LDA-NF
have a repulsion inside the nuclear surface and an attraction outside 
the nucleus, which are considerably different from each other in terms of 
the repulsion at $r \lesssim$ $R=$ 3.34 fm and the attractive pocket outside there;  
the potentials for LDA-S3, WS-sh and $t_\text{eff} \rho$ have an attraction 
at the nuclear center.
It was shown that the former potentials (a - c) were favored by the analysis of 
the ($\pi^-$, $K^+$) reaction,  rather than the latter ones (d - f), as discussed 
in Ref.~\cite{Harada05}. 
However, the radial distribution of the potential inside the nucleus 
and its strength at the center were hardly determined by fits to 
the ($\pi^-$, $K^+$) spectrum \cite{Harada05}.

\Figuretable{FIG. 1}%

It is important to investigate the $\Sigma$-nucleus potential for  
neutron-excess nuclei like ${^{208}{\rm Pb}}$, 
because one expects to obtain valuable information 
on the isovector component $U_1^\Sigma$ in the potential, 
while the $\Sigma$-nucleus potential for $^{28}$Si gives us 
information on the isoscalar component $U_0^\Sigma$.
In Fig.~\ref{fig:2}, we display the real and imaginary parts of 
several $\Sigma$-nucleus potentials for $^{208}$Pb, which is determined 
by fits to the $\Sigma^-$ atomic X-ray data. 
The potentials for DD-A$'$, DD-OBE and LDA-NF
have a strong repulsion inside the nuclear surface and 
an attraction outside the nucleus with a sizable absorption \cite{Batty94,Harada06}, 
and the potentials for LDA-S3 and $t_\text{eff} \rho$ have an attraction at the nuclear center.
In a previous paper \cite{Harada06}, we have shown that the former potentials fully 
reproduce the spectrum of the $^{209}$Bi($\pi^-$,$K^+$) reaction, 
rather than the latter ones.
Thus we have concluded that they provide the ability to explain 
the data of  the ($\pi^-$,$K^+$) reactions as well as those of the $\Sigma^-$ atoms; 
but it was impossible to discriminate among the radial distributions of the potentials 
for DD-A$'$, DD-OBE and LDA-NF inside the nucleus, 
and it was difficult to clearly see the contributions of the 
isoscalar and isovector components in this analysis \cite{Harada06}.

\Figuretable{FIG. 2}

\section{Theory}
\label{sect:calculation}

We calculate the differential cross sections by solving 
the radial part of the non-relativistic Schr{\"o}dinger equation
as a scattering problem: 
\begin{eqnarray}
\left[{\hbar^2 \over 2\mu}
\left(-{d^2 \over dr^2}+ {L(L+1) \over r^2} \right) 
+ U_{\Sigma}(r) + U_{\rm Coul}(r)\right] R_{L}(r)
= E R_{L}(r),
\label{eqn:e2}
\end{eqnarray}
where 
$R_L$ is a radial wave function with angular momentum $L$; 
$U_{\Sigma}$ is the $\Sigma$-nucleus potential, 
$U_{\rm Coul}$ is the Coulomb potential with a uniform distribution 
of charge for $R_{\rm C}=1.2A^{1/3}$,
$\mu$ is the $\Sigma^-$ nucleus reduced mass, 
and $E={\hbar^2k^2/(2\mu})$ is the incident energy of the center-of-mass frame. 
It is noted that the potentials of $U_{\Sigma}$ seem to include effects of 
a nuclear spin-orbit potential because they can reproduce the data of the $\Sigma^-$ 
atomic ($n\ell$) states and ($\pi^-$, $K^+$) reactions.
However, as far as the elastic scattering of $\Sigma^-$ hyperons at the low-energy 
like $E_{\rm lab}=$ 50 MeV is concerned, 
the effects of the spin-orbit potential on the differential cross section are 
negligible, as we will discuss later.

The angular distribution of the differential 
cross section in the elastic scattering is written as
\begin{eqnarray}
\sigma_{\rm el}(\theta)=|f(\theta)|^2
=|f_{\rm N}(\theta)+f_{\rm F}(\theta)|^2,
\label{eqn:e3}
\end{eqnarray}
where $f(\theta)$ is the elastic scattering amplitude,
which is often decomposed into traveling-waves decomposition 
of the nearside (N) and farside (F) components, 
$f_{\rm N}(\theta)$ and $f_{\rm F}(\theta)$ \cite{Fuller75,Mcvoy84}. 
They denote the sum of the Coulomb and nuclear parts as 
\begin{eqnarray}
f_{\rm N,F}(\theta)=f^{\rm (Coul)}_{\rm N,F}(\theta)
                   +f^{\rm (Nucl)}_{\rm N,F}(\theta)
\label{eqn:e4}
\end{eqnarray}
with 
\begin{eqnarray}
f^{\rm (Nucl)}_{\rm N,F}(\theta)
={i \over 2k}\sum_L (2L+1)e^{2i\sigma^{({\rm C})}_L}
(1-S_L)\tilde{Q}_L^{(\mp)}(\cos{\theta}), 
\label{eqn:e5}
\end{eqnarray}
where $\sigma^{({\rm C})}_L$ and $S_L$ are the Coulomb phase shift
and the S-matrix element in the elastic scattering, 
respectively. The traveling-wave function  
$\tilde{Q}_L^{(-)}$ ($\tilde{Q}_L^{(+)}$) corresponds to 
the nearside (farside) component, and it is defined in terms of 
the Legendre functions \cite{Fuller75}:
\begin{eqnarray}
\tilde{Q}_L^{(\pm)}(\cos{\theta})=
{1 \over 2}\left[ P_L(\cos{\theta})
\mp i {2 \over \pi}Q_L(\cos{\theta})
\right].
\label{eqn:e6}
\end{eqnarray}
In the semi-classical limit, they can be associated 
with trajectories that pass the near side 
and the far side of the scattering center.
Therefore, this decomposition gives a good 
understanding of the behavior of the angular distribution
depending on the nuclear potential that has an attraction 
and/or a repulsion with a strong absorption \cite{Fuller75,Mcvoy84}.
The angular distributions for the nearside and farside 
contributions denote 
$\sigma_{\rm N}(\theta)=|f_{\rm N}(\theta)|^2$ and 
$\sigma_{\rm F}(\theta)=|f_{\rm F}(\theta)|^2$, 
respectively.
Because the $\Sigma$-nucleus potential has a sizable absorption, 
the angular distribution of $\Sigma^-$ hyperons from nuclei
may behave similar to that of light composite nuclei rather than that of protons.

\section{Results and discussion}
\label{sect:results}

\begin{figure}[htb]
  \begin{center}
  \includegraphics[width=0.6\linewidth]{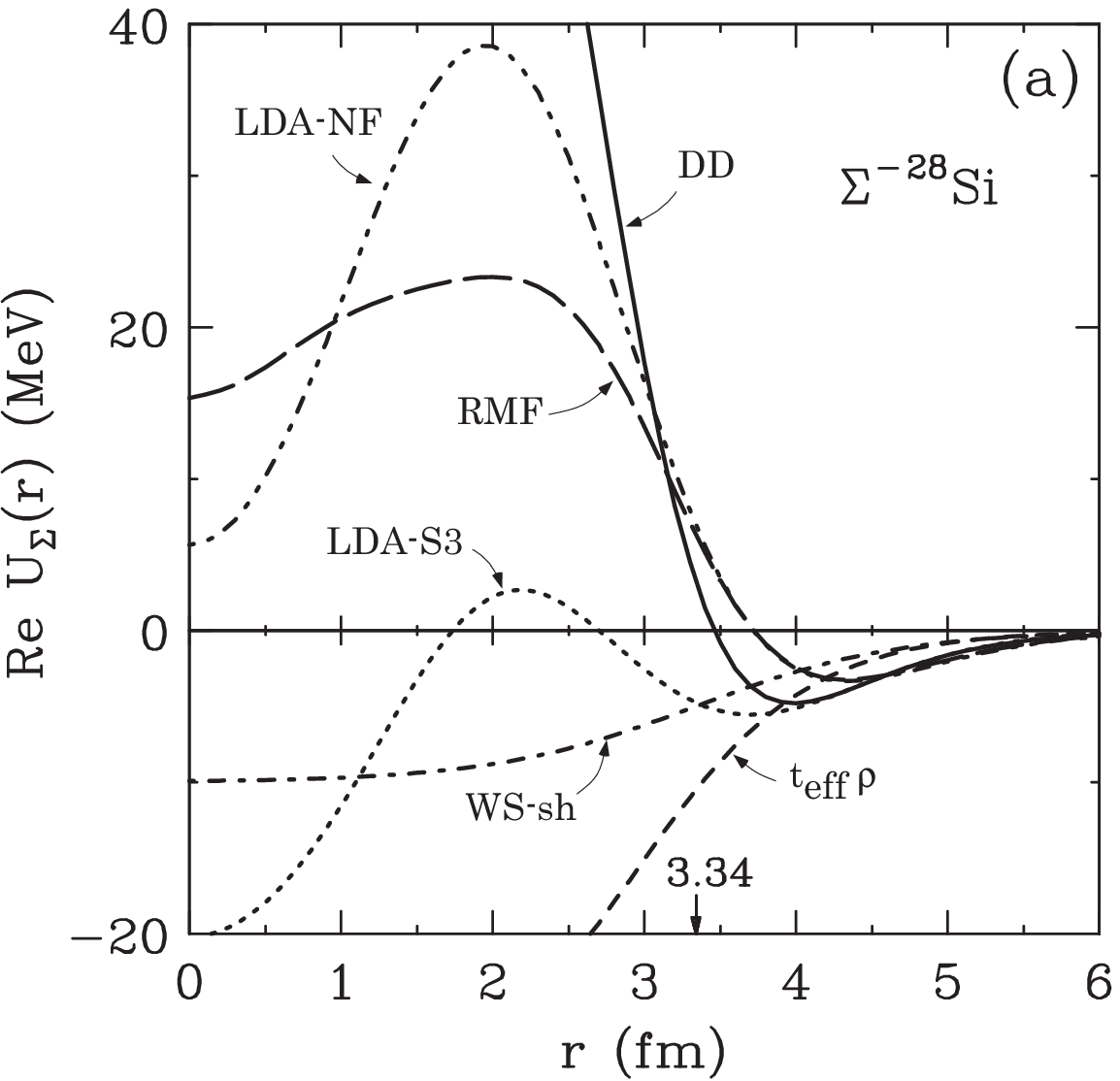}
  \includegraphics[width=0.6\linewidth]{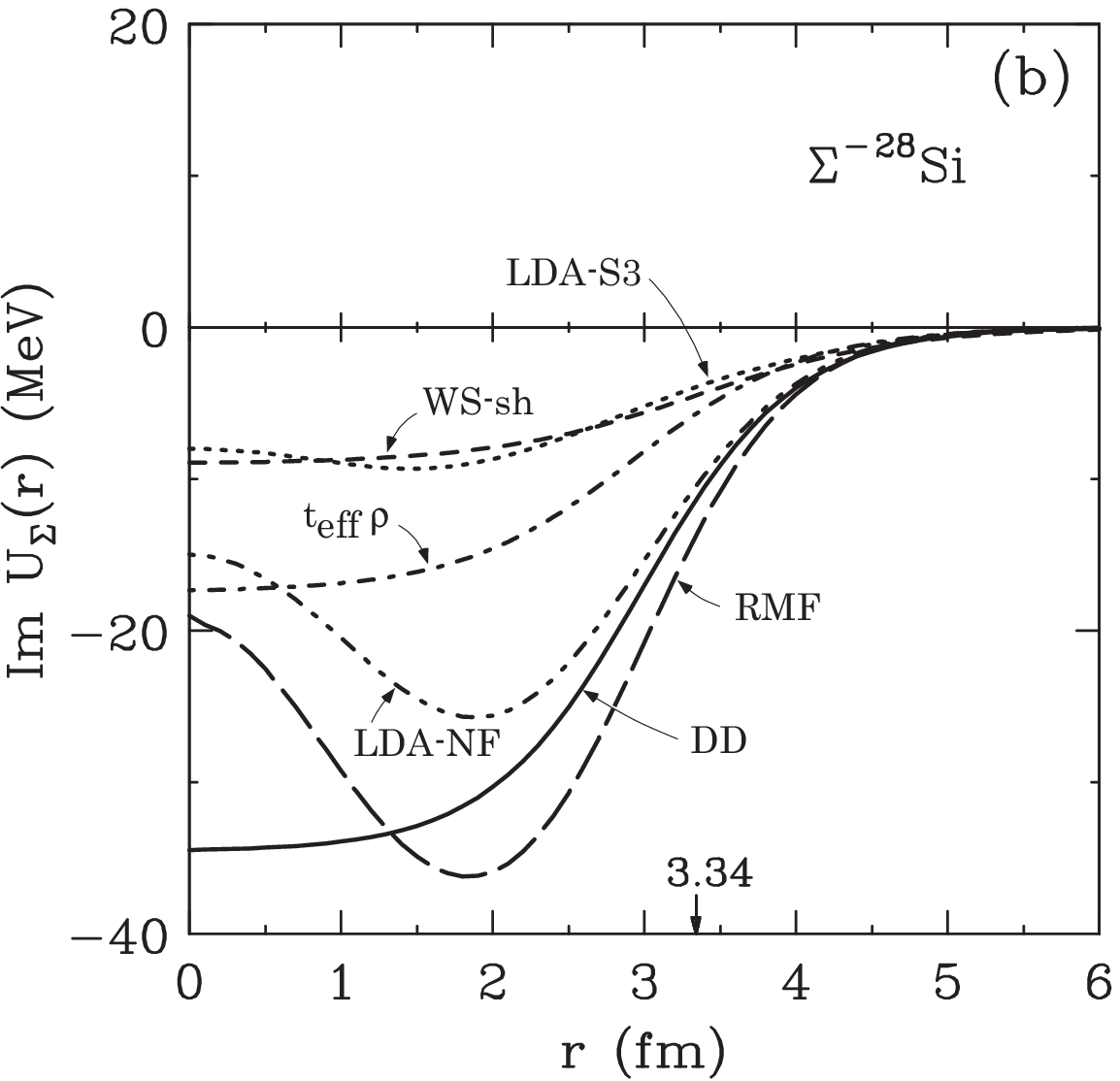}
  \end{center}
  \caption{\label{fig:1}
   Real (a) and  imaginary (b) parts of the $\Sigma$-nucleus potential 
   $U_{\Sigma}$ for $^{28}$Si,  
  as a function of the radial distance between the $\Sigma^-$ and the nucleus $^{28}$Si.
  The Coulomb potential is not included in the real part of each potential.
  Curves denote the potentials for DD, RMF, LDA-NF, 
  LDA-S3, WS-sh and $t_{\rm eff} \rho$ \cite{Harada05}.
  The arrows at $r= 1.1 A_{\rm core}^{1/3}$= 3.34 fm denote the nuclear radius of 
  the $\Sigma^-$-$^{28}$Si system. 
  }
\end{figure}

\begin{figure}[htb]
  \begin{center}
  \includegraphics[width=0.6\linewidth]{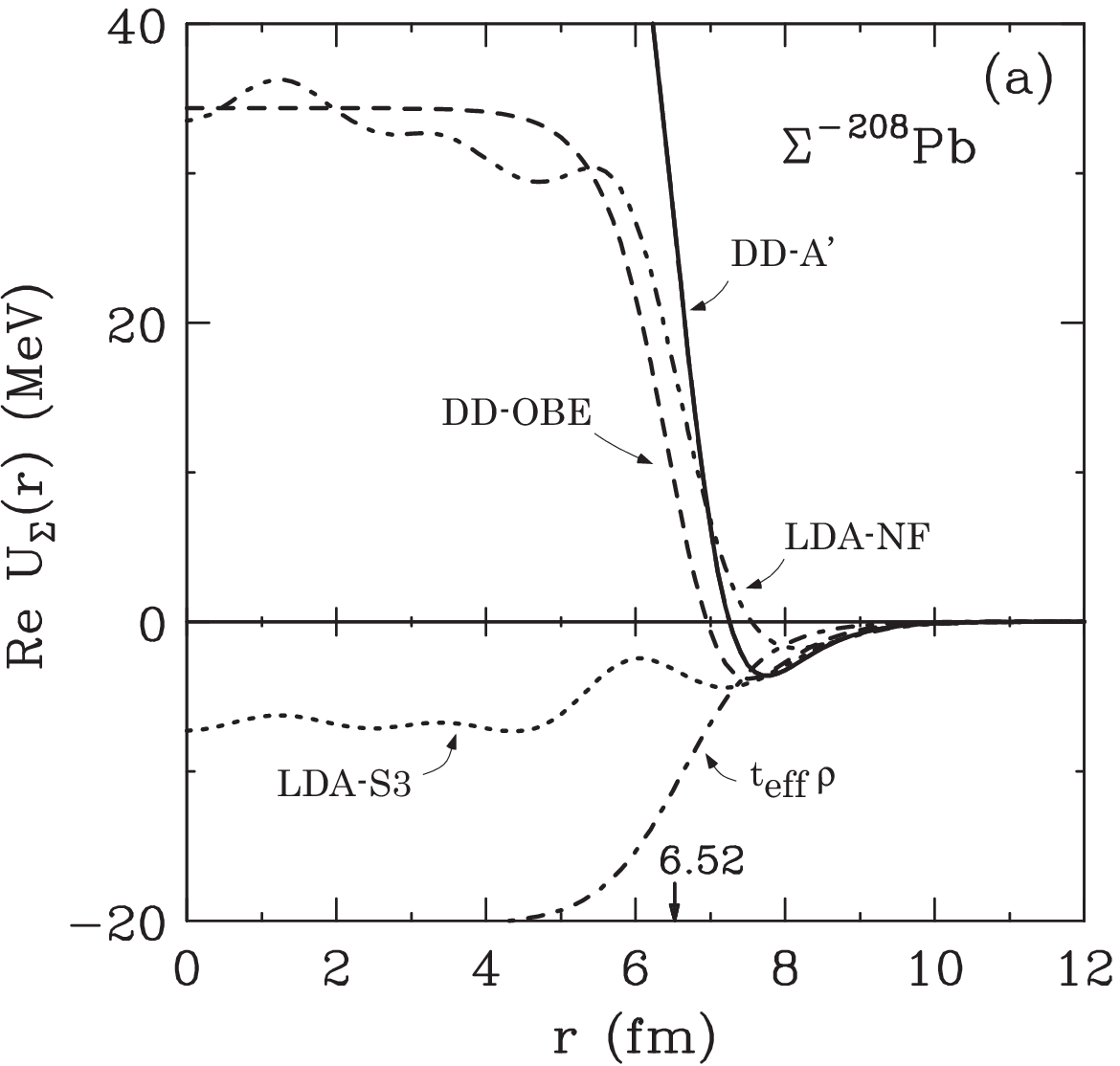}
  \includegraphics[width=0.6\linewidth]{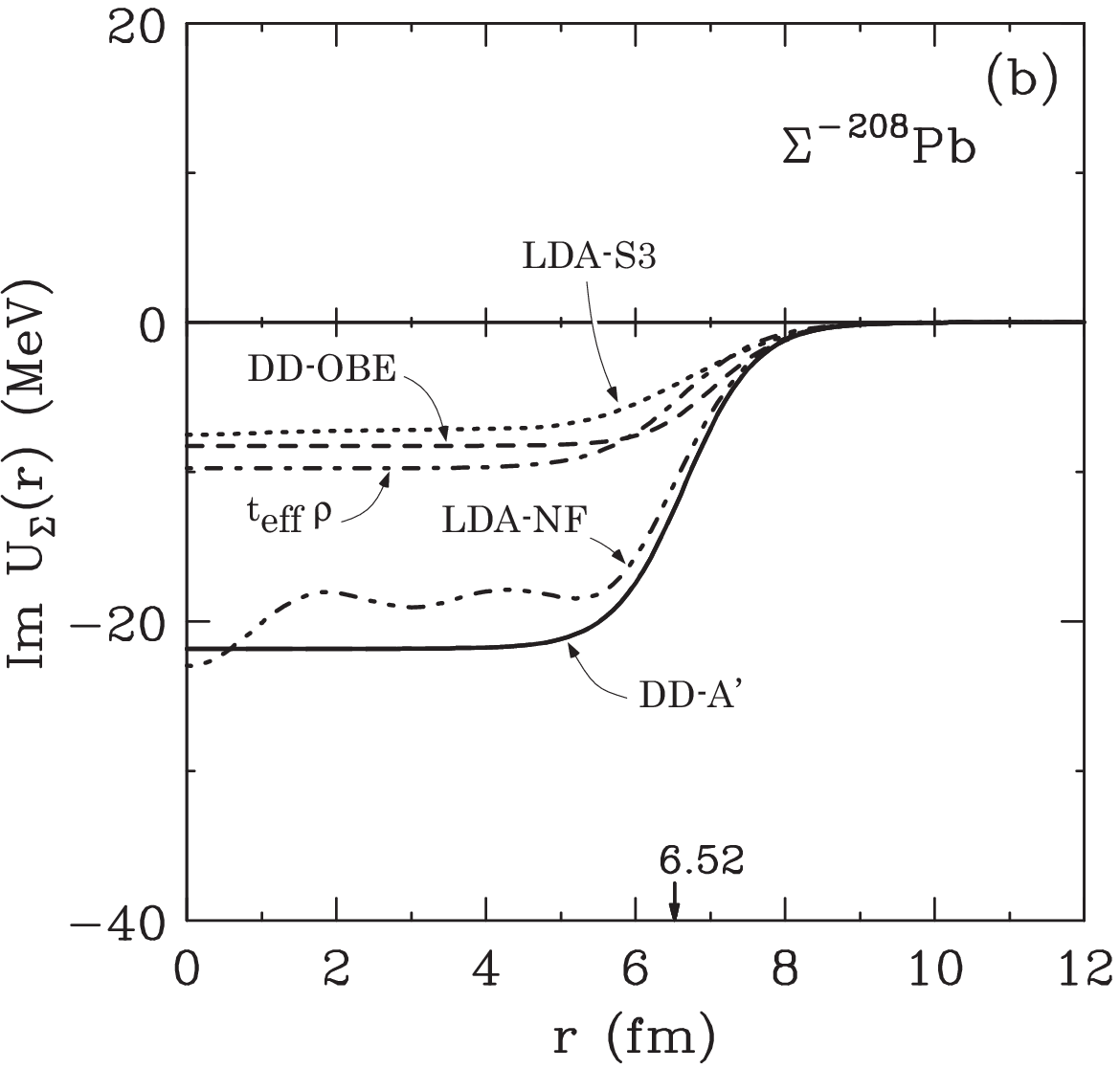}
  \end{center}
  \caption{\label{fig:2}
    Real (a) and  imaginary (b)  parts of the $\Sigma$-nucleus potentials 
   $U_{\Sigma}$ for $^{208}$Pb, 
   as a function of the radial distance between the $\Sigma^-$ and 
   the nucleus $^{208}$Pb.
   The Coulomb potential is not included in the real part of each potential.
   Curves denote the potentials for DD-A$'$, DD-OBE, LDA-NF, 
  LDA-S3 and $t_{\rm eff} \rho$ \cite{Harada06}.
  The arrows at $r= 1.1 A_{\rm core}^{1/3}$= 6.52 fm denote the nuclear radius of 
  the $\Sigma^-$-$^{208}$Pb system.
  }
\end{figure}

Let us consider the elastic scattering of $\Sigma^-$ hyperons
from $^{28}$Si \cite{Mares95,Furumoto10}. 
We calculate the angular distribution of the differential
cross section in this elastic scattering,  using several types of the 
$\Sigma$-nucleus potential (see Fig.~\ref{fig:1}).
Here we assume the $\Sigma^-$ incident energy of 
$E_{\rm lab}=$ 50 MeV in the laboratory frame. 

\begin{figure}[htb]
  \vspace{5mm}
  \begin{center}
  \includegraphics[width=1.0\linewidth]{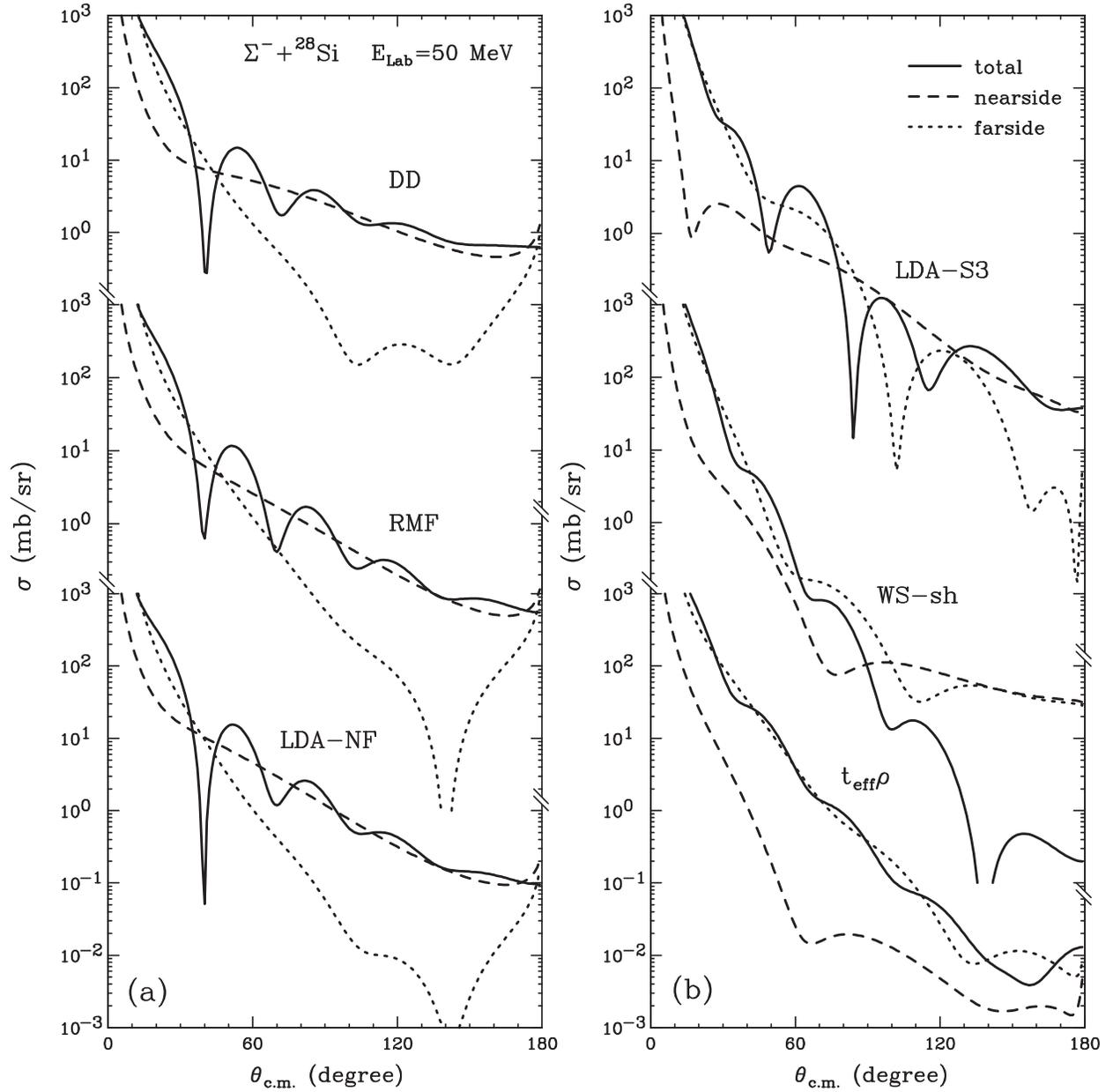}
  \end{center}
  \caption{\label{fig:3}
  Calculated angular distributions for the 50-MeV elastic scattering 
  of $\Sigma^-$ hyperons from $^{28}$Si. 
  Curves draw the absolute values obtained with the $\Sigma$-$^{28}$Si 
  potentials (a) for DD, RMF and LDA-NF, and 
  (b) for LDA-S3, WS-sh and $t_{\rm eff} \rho$.
  Solid, dashed and dotted curves denote the values for total, nearside and farside 
  components in the cross sections, respectively.
  }
\end{figure}

In Fig.~\ref{fig:3}, we show the calculated angular distributions of the cross section 
$\sigma_{\rm el}$ from $^{28}$Si, together with the nearside $\sigma_{\rm N}$ 
and farside $\sigma_{\rm F}$ components.
It is clearly seen that the angular distributions for DD, RMF and LDA-NF
differ from those for LDA-S3, WS-sh and $t_{\rm eff}\rho$.
This implies that the potentials of the former 
are fully distinguishable from those of the latter, and it supports 
our previous results on the analysis of the ($\pi^-$, $K^+$) reaction 
\cite{Harada05,Harada06}. 

In the angular distributions for DD, RMF and LDA-NF, 
the diffraction oscillations arise from 
the Fraunhofer interference between the nearside $f_{\rm N}$ and farside 
$f_{\rm F}$ amplitudes,
and reach their maximum amplitude at $\theta \simeq \bar{\theta}$, where 
the angle $\bar{\theta}$ for $\sigma_{\rm N}(\bar{\theta})=\sigma_{\rm F}(\bar{\theta})$
is called  a ``Fraunhofer crossover'' 
and hence $\bar{\theta} =$ 40$^\circ$$-$50$^\circ$ in the cases of these potentials. 
Forward of the crossover,  the farside is dominant, 
while for angles somewhat larger than $\bar{\theta}$, 
the nearside dominates.
This oscillations is damped with increasing $\theta$ 
because of the falloff of the farside components $\sigma_{\rm F}$.
The dominance of $\sigma_{\rm F}$ at small angles is caused by 
the Coulomb attraction for $\Sigma^-$ hyperons that have a 
negative charge, and by the attraction pocket of the $\Sigma$-nucleus 
potentials at the nuclear surface;  the dominance of $\sigma_{\rm N}$
at large angles is due to the strong repulsive components 
in the potentials for DD, RMF and LDA-NF.
It is very interesting because this situation is completely opposite
to that of the normal nucleus-nucleus scattering where $\sigma_{\rm N}$ 
dominates at small angles owing to the Coulomb repulsion, 
and $\sigma_{\rm F}$ becomes dominant with increasing $\theta$ 
owing to the strong attraction in nucleus-nucleus potentials
\cite{Fuller75,Mcvoy84}.
The behavior of the angular distribution in the nucleus-nucleus elastic scattering 
has been studied in the case of the repulsive nucleus-nucleus potential 
at high incident energies of $E/A \gtrsim$ 300 MeV \cite{Furumoto10a}.

In the cases of the potential for
$t_{\rm eff}\rho$, WS-sh and LDA-S3, 
the slope and oscillation pattern of their angular distributions 
differ appreciably.
For $t_{\rm eff}\rho$, 
we find that $\sigma_{\rm F}$ is dominant over all angles 
because the $\Sigma$-nucleus potential is attractive, 
in addition to the Coulomb attraction. 
For LDA-S3,  the oscillations are clearly observed owing to 
the Fraunhofer crossover between $\sigma_{\rm F}$ 
and $\sigma_{\rm N}$ at $\theta \simeq$ 90$^\circ$
and the magnitude of $\sigma_{\rm N}$ is something large at 
$\theta \gtrsim$ 90$^\circ$
because the real part of the potential is repulsive at 
$r \simeq$ 1.8$-$2.6 fm, as seen in Fig.~\ref{fig:1}.
The situation for WS-sh seems to be on the way from LDA-S3 
to $t_{\rm eff}\rho$.

Therefore, 
we show that the angular distribution of the 
$\Sigma^-$ elastic-scattering differential cross section from 
$^{28}$Si provides  
a classification for properties of the $\Sigma$-nucleus potentials.

\Figuretable{FIG. 3}

\begin{figure}[htb]
  \vspace{5mm}
  \begin{center}
  \includegraphics[width=1.0\linewidth]{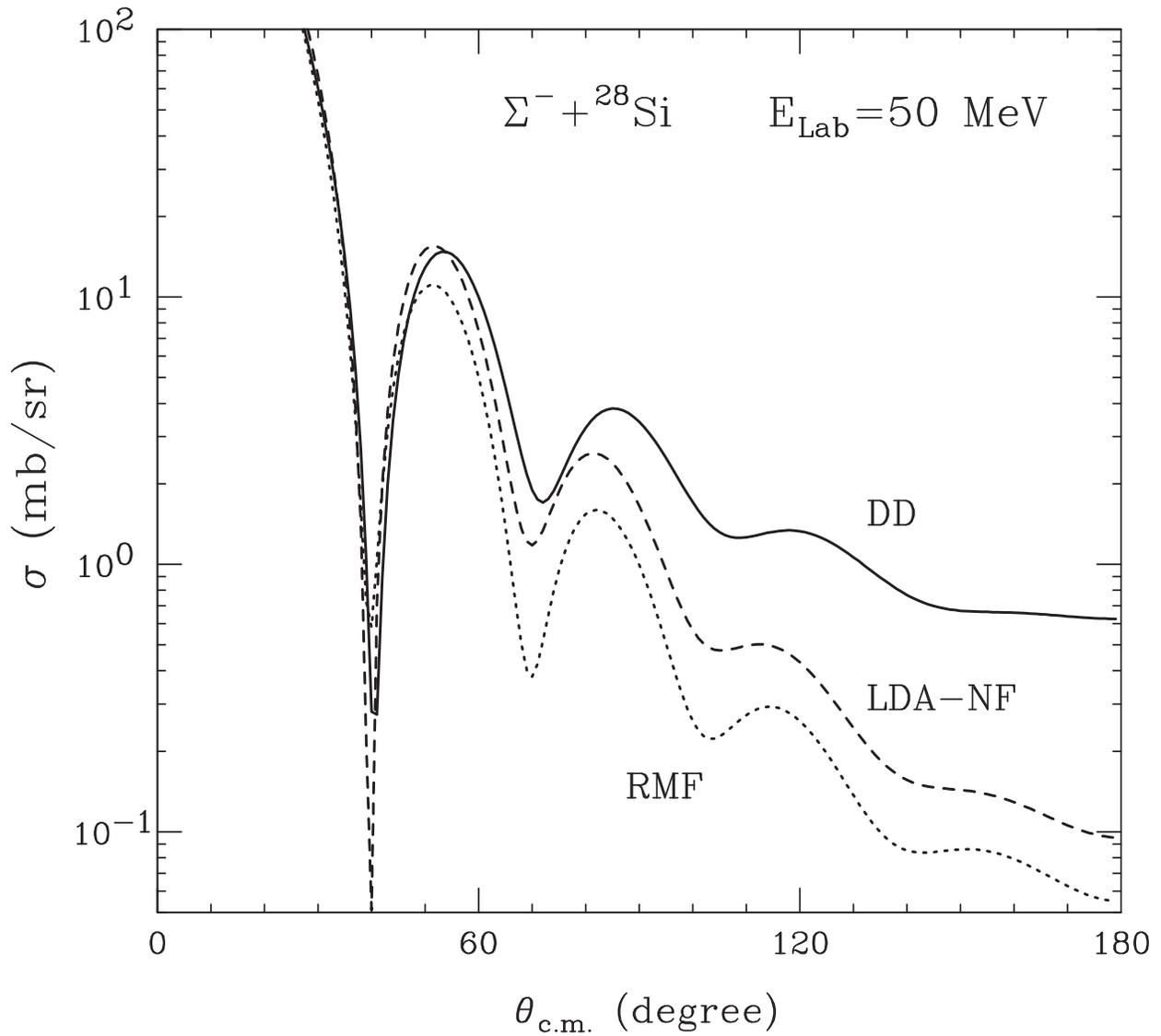}
  \end{center}
  \caption{\label{fig:4}
  A comparison with the calculated angular distributions 
  for the 50-MeV $\Sigma^-$ hyperon elastic scattering 
  from $^{28}$Si.
  Solid, dashed and dotted curves denote the values obtained with  
  the potentials for DD, RMF and LDA-NF in $\Sigma^-$-$^{28}$Si systems, 
  respectively. 
  }
\end{figure}

In Fig.~\ref{fig:4}, we compare the cross sections for DD, RMF and LDA-NF
in order to evaluate the detailed discrimination of properties 
of the potential inside the nucleus.
We notice that the magnitudes of their angular distributions ($\sigma_{\rm el}$) 
differ appreciably at angles $\theta \gtrsim 60^{\circ}$, 
where the magnitude and shape of $\sigma_{\rm el}$ are affected by 
the radial distribution of the potential inside the nuclear surface.
Indeed, a notch test suggests that the magnitude and shape of 
$\sigma_{\rm el}$ at angles $\theta \gtrsim 90^{\circ}$ are sensitive to the radial 
distribution of the potential at $r \simeq$ 2.6 fm, 
which corresponds to the region of  the inner repulsion 
of the potential, depending on the strength of the imaginary parts.

\Figuretable{FIG. 4}

\begin{figure}[htb]
  \vspace{5mm}
  \begin{center}
  \includegraphics[width=0.7\linewidth]{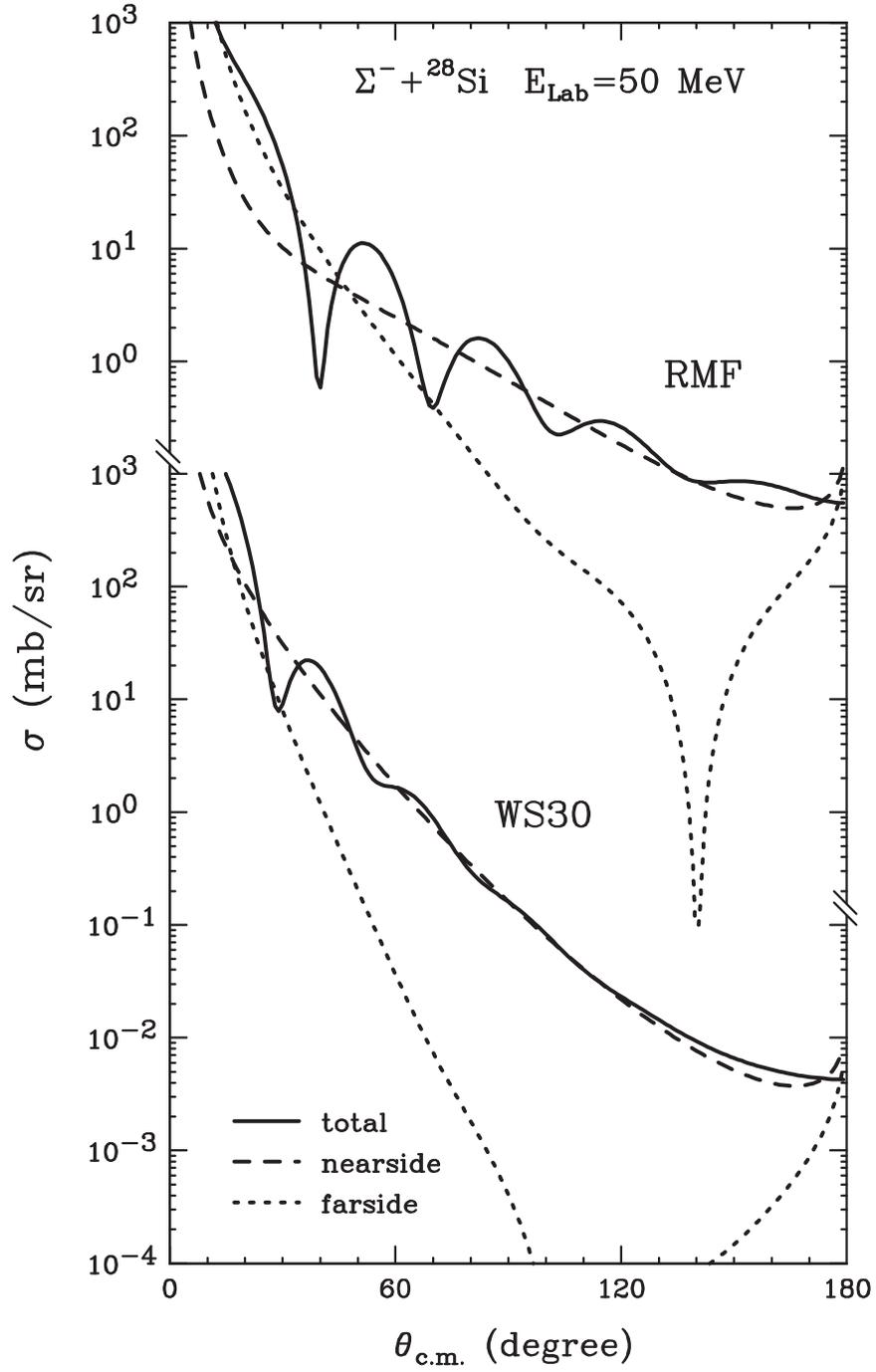}
  \end{center}
  \caption{\label{fig:5}
  Calculated angular distributions 
  for the 50-MeV $\Sigma^-$ hyperon elastic scattering 
  from $^{28}$Si, in comparison between RMF and WS30. 
  Solid, dashed and dotted curves denote the values for total, 
  nearside and farside components in the cross sections, respectively.
  }
\end{figure}

For DD, RMF or LDA-NF, moreover, the attractive pocket of the 
potential at the nuclear surface with the Coulomb attraction 
plays an important role in making a diffraction structure 
of the angular distribution, so that it leads the nearside component to be large, 
and causes a strong oscillation of the angular distribution.
To clarify this effect, 
we study behavior of the angular distribution $\sigma_{\rm el}$ 
using the WS potential that has only 
the repulsion in the real part and a sizable absorption 
in the imaginary part, for example, WS30 defined as
($V^\Sigma_0$, $W^\Sigma_0$)=($+30$ MeV, $-40$ MeV) in Eq.~(\ref{eqn:e1}).
In Fig.~\ref{fig:5}, we show the calculated angular distribution for WS30, 
as compared with that for RMF. 
We find that the magnitude of $\sigma_{\rm F}$ for WS30 
falls off rapidly on its steep slope with increasing $\theta$ from the 
forward angle. 
Thus this diffraction oscillations are shifted toward the forward angle and 
their magnitude becomes small. 
This is caused by the lack of the attraction at the nuclear surface 
in the repulsive WS30 potential. 

Consequently, it is shown that the angular distribution for the elastic 
scattering of 50-MeV $\Sigma^-$ hyperons from $^{28}$Si 
gives additional information to discriminate among the radial 
distributions of the potentials inside the nucleus, e.g.,  
inner repulsion and attractive pocket, 
which were not able to be identified by the analysis of the $\Sigma^-$ 
atomic X-ray and ($\pi^-$,$K^+$) reaction. 
It implies that the elastic scattering of $\Sigma^-$ hyperons
from nuclei is a powerful tool for identifying the radial distribution 
of the potential by the use of the diffraction pattern 
influenced by the nuclear repulsion and the Coulomb attraction, 
rather than $\Sigma^+$ hyperons acting on the nuclear repulsion 
with the Coulomb repulsion.

\Figuretable{FIG. 5}

\begin{figure}[htb]
  \vspace{5mm}
  \begin{center}
  \includegraphics[width=1.0\linewidth]{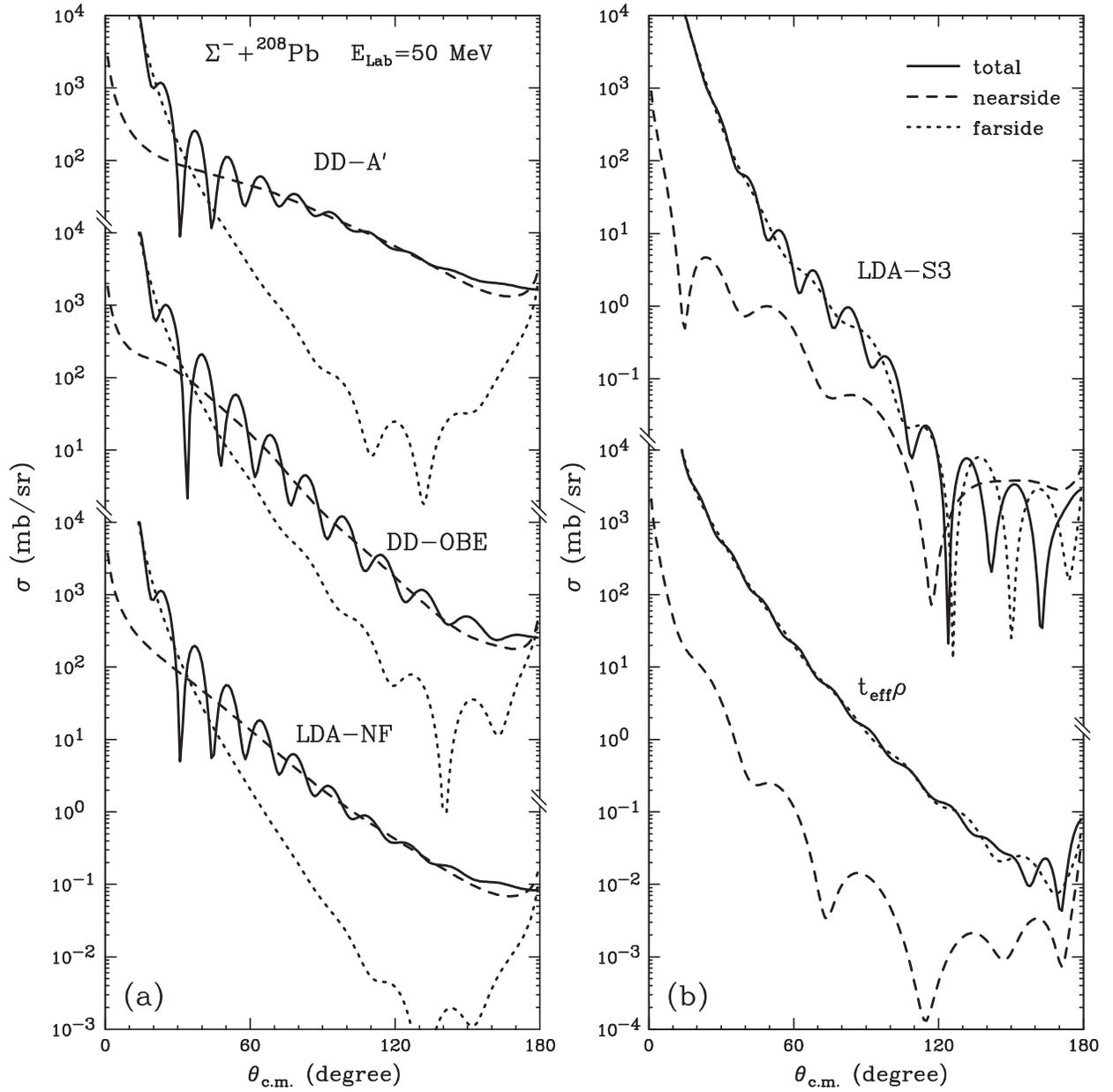}
  \end{center}
  \caption{\label{fig:6}
  Calculated angular distributions for the 50-MeV elastic scattering
  of $\Sigma^-$ hyperons from $^{208}$Pb.
  Curves draw the absolute values obtained with the $\Sigma$-$^{208}$Pb 
  potentials (a) for DD-A$'$, DD-OBE and LDA-NF, 
  and (b) for LDA-S3 and $t_{\rm eff} \rho$.
  See also the caption to Fig.~\ref{fig:3}.
  }
\end{figure}

We consider the elastic scattering of $\Sigma^-$ hyperons
from $^{208}$Pb, which has a large attraction of the Coulomb interaction. 
We calculate the angular distribution of the differential
cross section ($\sigma_{\rm el}$)  at $E_{\rm lab}=$ 50 MeV,  
using several types of the $\Sigma$-nucleus potential (see Fig.~\ref{fig:2}).
In Fig.~\ref{fig:6}, we show the calculated angular distributions of 
$\sigma_{\rm el}$,  together with the nearside 
$\sigma_{\rm N}$ and farside $\sigma_{\rm F}$ components, respectively. 

In the cases of DD-A$'$, DD-OBE and LDA-NF, 
we find the similar behavior of their results in the angular distribution, 
where a difference between the $\Sigma$-nucleus potentials for $^{208}$Pb
is not so enhanced, in comparison with that for $^{28}$Si. 
This recalls the fact that the effect on the $\Sigma$-nucleus 
potential is rather masked by the strong Coulomb potential 
in $^{208}$Pb, as discussed in Ref.~\cite{Harada09}.
However, we recognize that the diffraction pattern in $^{208}$Pb 
differs from that in $^{28}$Si as follows: 
As increasing $\theta$ the former $\sigma_{\rm F}$ falls off on a steep slope
more rapidly than the latter $\sigma_{\rm F}$, 
and hence the former $\sigma_{\rm N}$ dominates at $\theta \gtrsim$ 30$^\circ$. 
Thus the Fraunhofer crossover is slightly shifted 
toward the forward angle.
Moreover,  a lot of Fraunhofer oscillations in $^{208}$Pb appear, 
which may correspond to the grazing angular 
momentum $L_g \simeq$ 13 ($L_g \simeq$ 6 in $^{28}$Si) for LDA-NF
if $L_g$ is defined as a value at 
the transmission $T_L=$ 1/2 \cite{Mcvoy84}.
The oscillation spacing between maxima $\Delta \theta$ is also reduced 
in $^{208}$Pb. 

In the case of $t_{\rm eff}\rho$, we find that 
$\sigma_{\rm F}$ is dominant in all angles 
and $\sigma_{\rm N}$ is negligible,
so that the oscillations in $\sigma_{\rm el}$ are indistinctive. 
This originates from the fact that both the $\Sigma$-nucleus 
potential and the Coulomb potential are strongly attractive. 
For LDA-S3, we also find that $\sigma_{\rm F}$ is dominant as well as 
that of $t_{\rm eff}\rho$, but $\sigma_{\rm N}$ is not negligible 
because the potential for LDA-S3 has a weaker attraction than 
that for $t_{\rm eff}\rho$.  
The oscillations in $\sigma_{\rm el}$ are clearly shown
at angles $\theta >$ 120$^\circ$
by the appearance of the crossover.

\Figuretable{FIG. 6}

\begin{figure}[htb]
  \vspace{5mm}
  \begin{center}
  \includegraphics[width=1.0\linewidth]{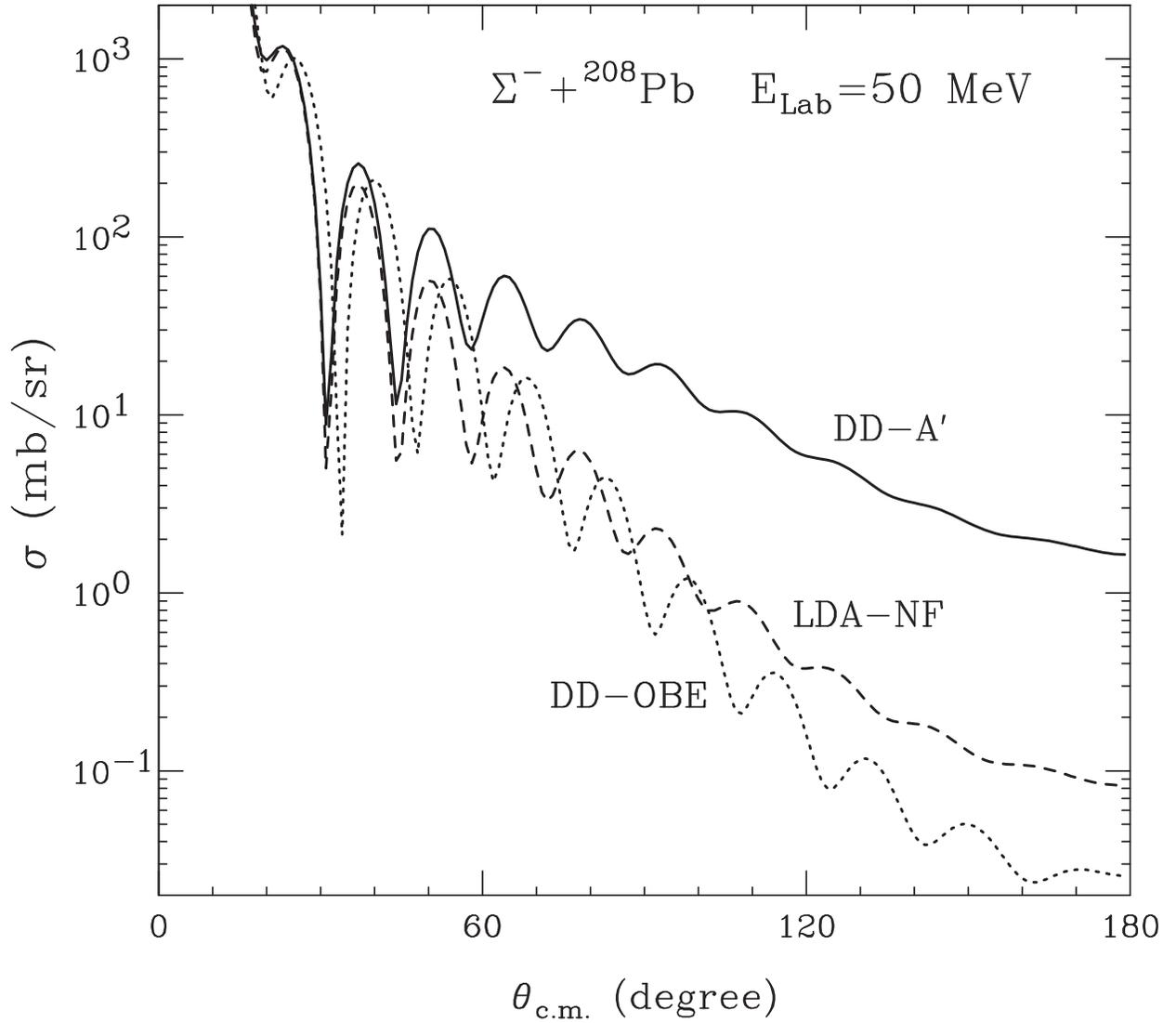}
  \end{center}
  \caption{\label{fig:7}
  A comparison with the calculated angular distributions 
  for the 50-MeV $\Sigma^-$ hyperon elastic scattering 
  from $^{208}$Pb. 
 Solid, dashed and dotted curves denote the values for 
 the potentials for DD-A$'$, LDA-NF and DD-OBE
 in $\Sigma^-$-$^{208}$Pb systems, respectively. 
  }
\end{figure}

In Fig.~\ref{fig:7}, we show the angular distributions of the cross sections 
for DD-A$'$, DD-OBE and LDA-NF
in order to evaluate the repulsive components
and radial distributions of their potentials inside the nucleus.
We compare the results on  DD-A$'$ with those on DD-OBE, 
because the former has the inner repulsion of $\sim$80 MeV 
at the nuclear center and the latter has that of $\sim$30 MeV
in the similar potential form, depending on the imaginary part 
of individual potentials.
Thus we realize that the magnitudes of their $\sigma_{\rm el}$ 
are sufficiently discriminable at $\theta \gtrsim$ 60$^\circ$.  
Moreover, let us compare the results on LDA-NF with those on DD-OBE,
because the radial distributions of their potentials differ markedly
whereas the inner repulsions at the nuclear center are almost the same.  
Thus we can see that the oscillation patterns of their $\sigma_{\rm el}$ 
are slightly different.
Consequently, we recognize that the angular distribution of the $\Sigma^-$ 
elastic-scattering differential cross section from $^{208}$Pb 
provides the ability to discriminate among the radial distributions 
of the potentials inside the nucleus.

\Figuretable{FIG. 7}

As mentioned in Sect.\ref{sect:calculation}, 
the potentials of $U_{\Sigma}$ seem to include the effects of the nuclear 
spin-orbit potential because they can reproduce the data of the 
$\Sigma^-$ atomic ($n\ell$) states and  ($\pi^-$, $K^+$) reactions,
though our calculations do not deal with the spin-orbit term explicitly.
However, it is one of the important subjects to study the spin-orbit 
potential for a $\Sigma$ hyperon \cite{Dover89}, whereas the experimental
information is extremely limited.
According to several theoretical predictions 
\cite{Morimatsu84,Dover84,Mares94,Kaiser07}, here, we consider the spin-orbit 
potential for $\Sigma$ 
with the strength of $V^{\Sigma}_{\rm so}\simeq$ ${1 \over 2}V^{N}_{\rm so}$,
where $V^{N}_{\rm so}$ for a nucleon.
When we use the potential for RMF including artificially a spin-orbit term of
$V^{\Sigma}_{\rm so}({1/r})[df(r)/dr]{\bm \sigma}{\cdot}{\bm L}$
which is often used, 
we show that the calculated angular distributions are almost the same as 
those for the original RMF in the elastic scattering of 50-MeV $\Sigma^-$
hyperons from $^{28}$Si, e.g., $\sigma_{\rm el}(\theta)$ at the first maximum
for $\theta \simeq$ 50$^\circ$ is increased by only about 6\%.
This small difference originates from the fact that the $\Sigma^-$ elastic 
scattering at $E_{\rm lab}=$ 50 MeV is regarded as a low-energy one, 
in comparison with high-energy scatterings of $E_{\rm lab} \simeq$ 100$-$300 MeV, 
where the spin-orbit effects could fairly act because the contribution of 
large $L$'s to the differential cross section is important.
For $^{208}$Pb the situation is the same as that for $^{28}$Si.

\medskip
\section{Conclusion}
\label{sect:summary}

We have theoretically investigated the elastic scattering of 50-MeV $\Sigma^-$ hyperons
from $^{28}$Si and $^{208}$Pb in order to clarify the radial distribution
of the $\Sigma$-nucleus potentials.
The angular distributions of their differential cross sections 
have been calculated using several potentials that can explain 
the experimental data of the $\Sigma^-$ atomic X-ray and ($\pi^-$, $K^+$) 
spectra simultaneously. 
We have discussed the behavior of the magnitude and oscillation pattern 
in the angular distribution by the use of the nearside/farside decomposition of 
the elastic scattering amplitude, and we have examined a competition between 
the attraction of the Coulomb interaction and the 
repulsion/attraction in the $\Sigma$-nucleus potentials.
As far as the low-energy elastic scattering like $E_{\rm lab}=$ 50 MeV is concerned, 
the effects of the spin-orbit potential on the angular distribution are very small.

In conclusion, the angular distribution of the differential cross sections in 
the 50-MeV $\Sigma^-$ elastic scattering from $^{28}$Si and $^{208}$Pb  
provides additional information to discriminate the nature of the repulsion/attraction inside the 
nuclear surface in the $\Sigma$-nucleus potentials where it was not uniquely 
determined by the $\Sigma^-$ atomic X-ray data and 
the ($\pi^-$, $K^+$) and ($K^-$, $\pi^+$) spectra. 
We expect that the elastic scattering experiments of $\Sigma^-$ hyperons from nuclear 
targets are curried out at J-PARC facilities in the future, 
in spite of some experimental difficulties.  
More theoretical investigations of the $\Sigma^-$ scattering 
for several incident energies and targets are required. 

\begin{acknowledgments}
The authors are obliged to Professor K.~Tanida and Dr. K.~Miwa 
for valuable discussion and comments.
This work was supported by Grants-in-Aid for Scientific Research
(C) (No.~22540294).
\end{acknowledgments}


\clearpage

\end{document}